%% file: gluino_analysis_prl_p14.tex
\documentclass[aps,prl,showpacs,twocolumn,groupedaddress]{revtex4}  
\usepackage{graphicx}  
\usepackage{dcolumn}   
\usepackage{bm}        
\usepackage{amssymb}   

\begin{document}
\input{definitions}


\hspace{5.2in}\mbox{FERMILAB-PUB-07-100-E} 

\title{Search for stopped gluinos from \ppbar\ collisions at \1960}
\input{list_of_authors_r2.tex} 

\date{May 2, 2007}

\begin{abstract}
Long-lived, heavy particles are predicted in a number of models
beyond the standard model of particle physics.  We present the first
direct search for such particles' decays, occurring up to 100 hours
after their production and not synchronized with an accelerator
bunch crossing. We apply the analysis to the gluino ($\tilde g$),
predicted in split supersymmetry, which after hadronization can
become charged and lose enough momentum through ionization to come
to rest in dense particle detectors. Approximately 410~\ipb\ of
\ppbar\ collisions at \1960\ collected with the D0 detector during
Run II of the Fermilab Tevatron collider are analyzed in search of
such ``stopped gluinos'' decaying into a gluon and a neutralino
($\tilde{\chi}_1^0$), reconstructed as a jet and missing energy. No
excess is observed above
background, and limits are placed on the (gluino cross section)
$\times$ (probability to stop) $\times$ [BR($\tilde
g$\rarrow$g\tilde{\chi}_1^0$)] as a function of the gluino and
$\tilde{\chi}_1^0$ masses, for gluino lifetimes from 30~$\mu$s --
100 hours.
\end{abstract}


\maketitle

Split supersymmetry is a relatively new variant of supersymmetry
(SUSY), in which the SUSY scalars are heavy
compared to the SUSY fermions \cite{Arkani-Hamed:2004yi}. Due to the
scalars' high masses, gluino decays are suppressed, and the gluino
can be long-lived. Other new models, such as Gauge-mediated SUSY,
can also predict a long-lived gluino or other heavy, colored,
long-lived particles \cite{Pape:2006ar}. The gluinos hadronize into
``R-hadrons" \cite{rhadron}, colorless bound states of a gluino and
other quarks or gluons.
As studied in Ref.~\cite{Arvanitaki:2005nq}, some 30\%
of R-hadrons at the Tevatron can become ``stopped gluinos" by
becoming charged through nuclear interactions, losing all of their
momentum through ionization, and coming to rest in surrounding dense
material. We present the first direct search for the decays of such
particles, with deposited hadronic energy not in-time with a \ppbar\
collision.

A data sample corresponding to an integrated luminosity of
410$\pm$25~\ipb\ \cite{lumi}, taken with the D0
detector~\cite{Abazov:2005pn} from November 2002 to August 2004, has
been analyzed to search for stopped gluinos.
The D0 detector has a magnetic central tracking system surrounded by
a uranium/liquid-argon calorimeter, contained within a muon
spectrometer. The tracking system, located within a 2~T solenoidal
magnet, is optimized for pseudorapidities $|\eta|<2.5$, where $\eta
= -\ln[\tan(\theta/2)]$, and $\theta$ is the polar angle with
respect to the proton beam direction ($z$). The calorimeter has a
central section (CC) covering up to $|\eta| \approx 1.1$, and two
end calorimeters (EC) extending coverage to $|\eta|\approx 4.2$, all
housed in separate cryostats~\cite{run1det}. The calorimeter is
divided into an electromagnetic part followed by fine and coarse
hadronic sections. Calorimeter cells are arranged in
pseudo-projective towers of size 0.1$\times$0.1 in $\eta\times\phi$,
where $\phi$ is the azimuthal angle. The muon system consists of a
layer of tracking detectors and scintillation trigger counters in
front of 1.8~T iron toroidal magnets (the A layer), followed by two
similar layers behind the toroids (the B and C layers), which
provide muon tracking for $|\eta|<2$. The luminosity is measured
using scintillator arrays located in front of the EC cryostats,
covering $2.7<|\eta|<4.4$. The trigger system comprises three levels
(L1, L2, and L3), each performing an increasingly detailed event
reconstruction in order to select the events of interest.

We search for stopped gluinos decaying into a gluon and a
neutralino, $\tilde{\chi}_1^0$. The analysis has slightly reduced
sensitivity for $\tilde g$\rarrow\qqbar$\tilde{\chi}_1^0$, which may
be a large fraction of the decays, depending on the SUSY parameters.
The gluino lifetime is assumed to be long enough such that the decay
event is closest in time to an accelerator bunch crossing later than
the one that produced the gluino. For the L1 trigger to be live
again during the decay even if the production event was triggered
on, this lifetime must be at least $30~\mu$s, due to trigger
electronics deadtime. The efficiency for recording the gluino decay
is modeled as a function of the gluino lifetime, up to 100 hours.
When the decay occurs during a bunch crossing with
no other inelastic \ppbar\ collision, the signal signature is a
largely empty event with a single large transverse energy (\et)
deposit in the calorimeter, reconstructed as a jet and large missing
transverse energy (\met).


The trigger for each event requires that neither of the luminosity
scintillator arrays fired. At least two calorimeter towers of size
$\eta\times\phi$=0.2$\times$0.2 with \et\gt 3 \gev\ are also
required at L1. Jets are reconstructed with the Run II Improved
Legacy Cone Algorithm \cite{RunIIcone} with a cone of radius 0.5 in
$\eta\times\phi$ space. A reconstructed jet with \et\gt 15 \gev\ is
required at L3.
Offline, we require exactly one jet in the event with $E$\gt 90
\gev, and no other jets with \et\gt 8 \gev. The calorimeter
requirements in the trigger are nearly 100\% efficient for events
that pass the 90 \gev\ offline threshold.

To simulate stopped gluino decays, the {\sc pythia} \cite{pythia}
event generator is used to produce $Z$+gluon events, with the $Z$
boson forced to decay to neutrinos. Initial-state radiation is
turned off, as are multiple parton interactions.
The spectator particles coming from the rest of the \ppbar\
interaction, such as the underlying event, are removed by removing
all far-forward particles with $|p_{z}/E|>0.95$. The location of the
interaction point is placed inside the calorimeter, and events are
further weighted such that the final decay position distribution is
that expected for stopped gluinos. The radial location of the gluino
when it decays depends on the way gluinos lose energy via ionization
and stop in the calorimeters. This calculation was performed
\cite{Arvanitaki:2005nq} for a distribution of material similar to
that of the D0 calorimeters and a gluino velocity distribution as
expected from production at the Tevatron. The $\eta$ distribution is
determined by the fact that gluinos would tend to be produced near
threshold at the Tevatron, and that only slow gluinos would stop.
The gluinos are thus expected to be distributed proportionally to
$\sin\theta$. More than 75\% of gluinos that stop have \aeta\lt 1.
Because the gluinos are at rest and with their spin randomly
oriented when they decay, the gluon is emitted in a random
direction. Thus a random 3D rotation is applied to the
simulated particles.

The energy of the gluon, which hadronizes and fragments into a jet,
depends on the gluino and neutralino masses: $E = (M_{\tilde
g}^2-M_{\tilde{\chi}_1^0}^2)/2 M_{\tilde g}$.
We generate four samples of stopped gluinos, containing about 1000
events each, using a {\sc geant}-based \cite{geant} detector
simulation and reconstructed using the same algorithms as data. They
correspond to gluino masses of 200, 300, 400, and 500 \gev, with a
neutralino mass of 90 \gev. These samples correspond to generated
gluon energies of 80, 137, 190, and 242 \gev, respectively.
Simulated jets are corrected for relative differences between the
data and simulation jet energy scales. The calorimeter electronics
sample the shaped ionization signal only once per bunch crossing, at
the assumed peak of the signal for jets originating from a \ppbar\
interaction, but the gluino decay can occur at any time with respect
to a bunch crossing. So jet energies in the simulation are also
corrected (downwards) according to a model of this ``out-of-time''
calorimeter response. The average degradation of energy is 30\%,
although more than half of the jets are not significantly degraded.


The primary source of background is cosmic muons, which are able to
fake a gluino signal if they initiate a high-energy shower within
the calorimeter. Hard bremsstrahlung is responsible for the majority
of the showers. These showers tend to be very short, since they are
electromagnetic in nature and thus have small lengths compared to
hadronic showers.
However, sometimes a wide, hadronic-like, shower can be created
either due to deep-inelastic muon scattering, fluctuations of the
shower, or detector effects.
Cosmic muons can usually be identified by the presence of a
reconstructed high-energy muon. A coincidence of muon hits in the B
and C layers of the muon system, behind the thick iron toroid
magnet, is very strong evidence of a muon. The A layer muon hits are
often also caused by the signal, due to particles escaping the
calorimeters, so are difficult to use for background rejection.
Sometimes the muon is not detected, due to detector inefficiencies,
being out-of-time with the bunch crossing, or the limited
acceptance.

Another source of background events is beam-halo muons, or
``beam-muons." These are muons, synchronized with the \ppbar\ bunch
crossings and traveling nearly parallel to the beam.
Often, one or more muon scintillator hits can be associated with the
muon, and the muon is measured to be within $\Delta t$\lt 10 ns of a
bunch crossing. Another feature of the beam-muons is that they are
nearly all in the plane of the accelerator beam.
Beam-muon showers are also typically very narrow in $\phi$, causing
this background to be negligible once wide calorimeter showers are
required.


Since the trigger requires no signal in the luminosity scintillator
arrays, nearly all of the \ppbar\ beam produced backgrounds are
eliminated. An exception is diffractive events with forward rapidity
gaps in both the positive and negative $\eta$ regions.
Typical \ppbar\ events have a primary vertex (PV) reconstructed from
tracks which originate near to each other along the beamline, where
the \ppbar\ interaction occurred. Dijet events in the same data
sample are studied to understand the \met\ spectrum and PV
reconstruction efficiency for beam-related backgrounds. After
requiring no PV to be reconstructed and large \met\ (implicit from
the requirement of a single high-energy jet), the \ppbar\ events are
negligible.

Other sources of physics background considered are cosmic neutrons
and neutrinos, both of which are found to be negligible. Cosmic
neutrons would have to penetrate the thick iron toroid. Those
neutrons that did reach the calorimeter would shower preferentially
in the outer layers on the top of the calorimeter, which is not
observed.

Finally, since the signal process is rare, we also consider
occasional fake signals caused by detector readout errors or
excessive noise. We require the jet to be in \aeta\lt 0.9, since the
forward regions of the calorimeter are observed to have more
frequent (yet still rare) problems. Also, the gluino signal tends to
be concentrated in the central detector region. Remaining problems
are isolated to a specific set of runs, detector region, or both,
and such events are removed.

\begin{table} \centering
\caption{The selections applied, and the number of events passing in
data and for a simulated signal with $M_{\tilde g}$=400 \gev\ and
$M_{\tilde{\chi}_1^0}$=90 \gev.}
\begin{tabular}{ccc}
  \hline
  \hline
  Selection & Data Events & Signal Events \\
  \hline
Total & 7199133 & 2000 \\
Exactly one jet (\et\gt 8 GeV) & 3691036 & 1678 \\
Jet \aeta\lt 0.9 & 2742353 & 1505 \\
Jet E\gt 90 GeV & 202568 & 805 \\
No PV & 198380 & 803 \\
Data quality & 189781 & 772 \\
Jet $\eta$ and $\phi$ widths \gt 0.08 & 5994 & 410 \\
Jet n90 \gt 10 & 1402 & 383 \\
No muons & 109 & 357 \\
  \hline
  \hline
\end{tabular}
\label{table:selections}
\end{table}

The following criteria are used to select events containing
``wide-showers'': jet $\eta$-width and $\phi$-width \gt 0.08 and jet
$n_{90}$ $\ge$10, where $n_{90}$ is the smallest number of
calorimeter towers in the jet that make up 90\% of the jet
transverse energy. The reverse criteria define a ``narrow-shower."
Criteria are also defined which select events containing ``no-muon''
or a ``cosmic-muon." An event contains no-muon if there are no B-C
layer muon segments in the event, and no A layer segments with
$\Delta\phi$\gt 1.5 radians from the jet direction. Cosmic-muon
events have at least one B-C layer muon segment with $|\Delta t|$\gt
10 ns from the bunch crossing time. A candidate stopped gluino decay
event contains both a wide-shower and no-muon.

\begin{figure}\centering
\includegraphics[width=1.75in]{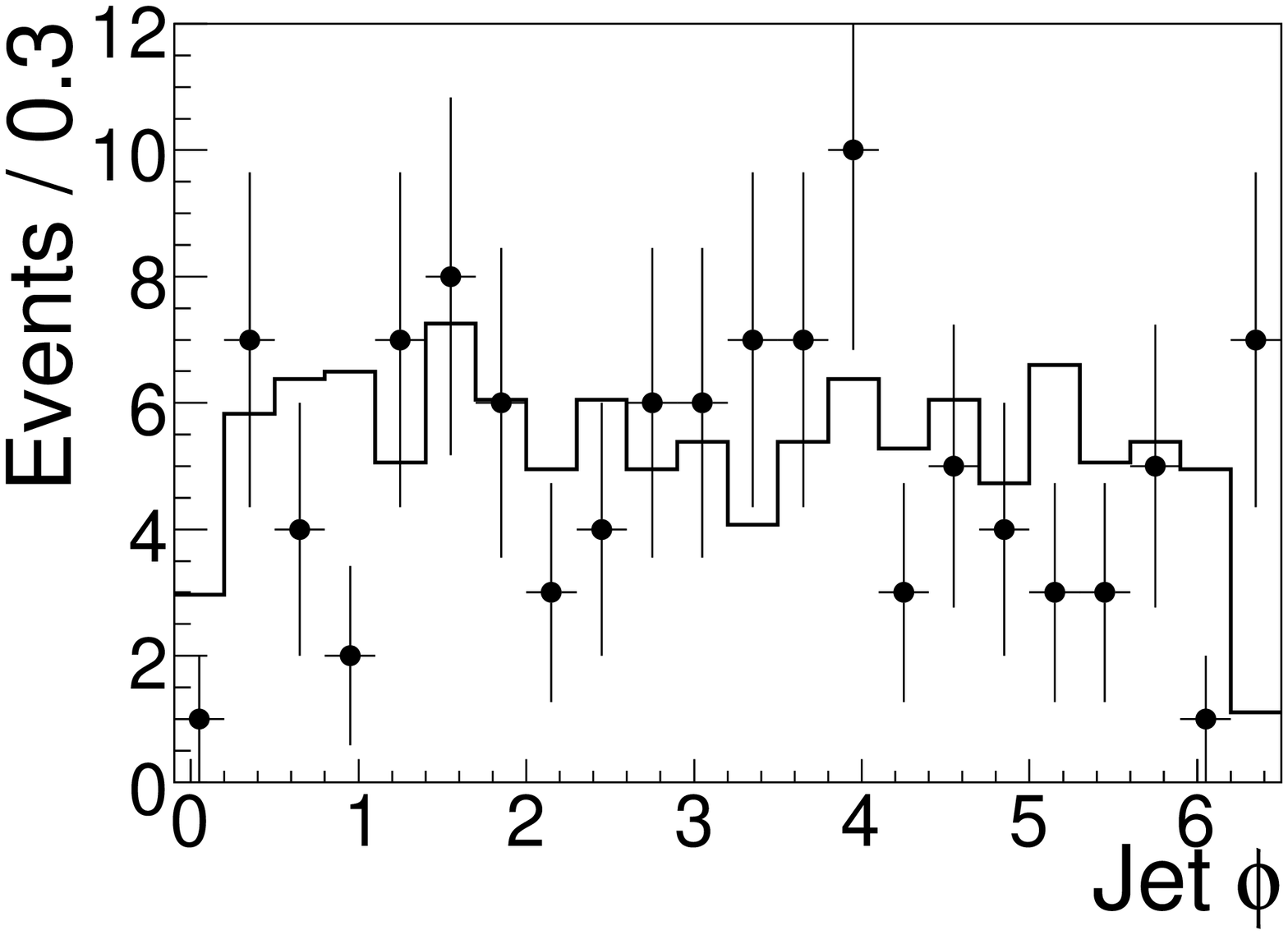}\includegraphics[width=1.75in]{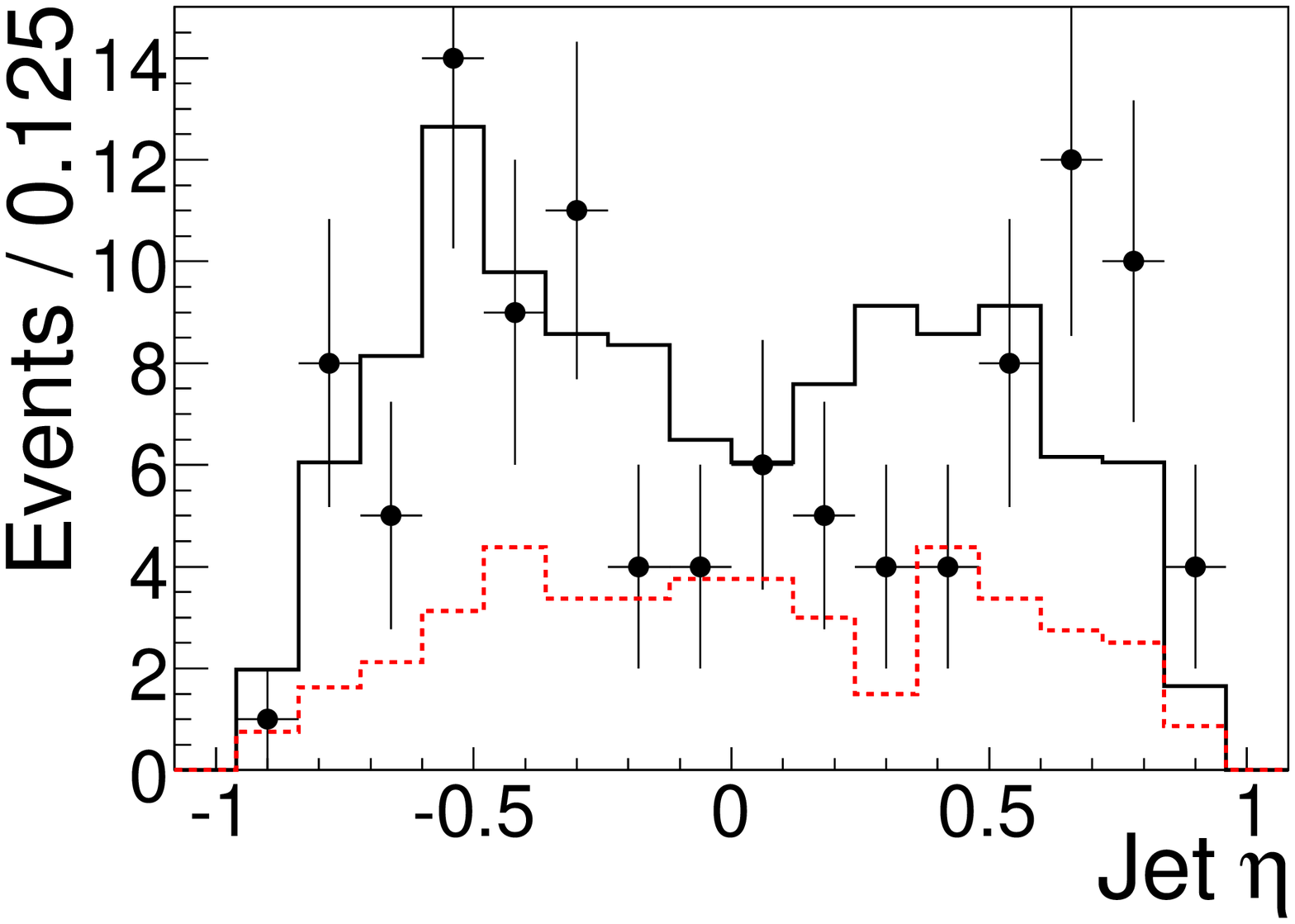}
\includegraphics[width=3.2in,height=1.8in]{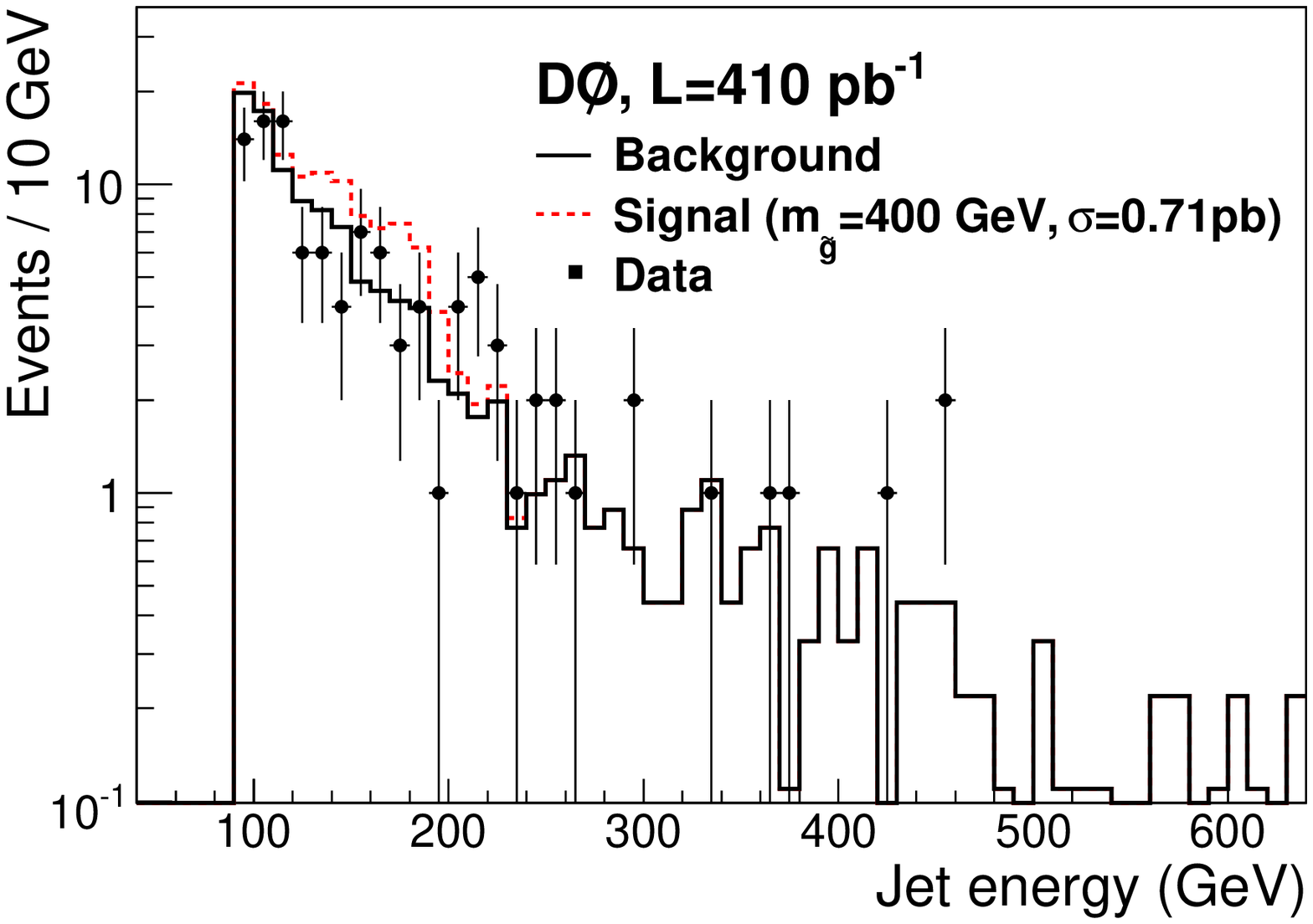}
\caption{A comparison of the wide-shower no-muon data (points) to
the expected background from cosmic muons (solid histogram) and a
simulated signal
(dashed histogram).} \label{fig:back_log}
\end{figure}

To estimate the number of such wide-shower no-muon events expected
from cosmic muon background, we use the assumption that the
probability not to reconstruct a cosmic muon in the muon system is
independent of whether the muon's shower in the calorimeter is
narrow or wide. A subset of the narrow-shower data sample is defined
which is nearly devoid of beam-muons by requiring a shower out of
the accelerator plane. This cosmic-muon narrow-shower data subset
has a similar $\eta$ distribution to the wide-shower data, and the
$\eta$ and $\phi$ shower width distributions are not altered
significantly when requiring a muon. The probability to not
reconstruct the muon in this narrow-shower data sample is measured
to be 0.11$\pm$0.01, independent of shower energy. This probability
is applied to the wide-shower cosmic-muon data sample to predict the
jet energy spectrum of wide-shower no-muon background events, as
shown in Fig.~\ref{fig:back_log}. The data agree with the estimated
background from cosmic muons. There is no significant excess in any
jet energy range, and the data has the predicted shape in $\eta$ and
$\phi$.

We search for a signal in jet energy ranges with widths chosen from
the jet energy resolutions of the simulated signal samples. The
ranges are from $M-\sigma/2$ to $M+2\sigma$, where $M$ is the mean
jet energy of the sample and $\sigma$ is the sample's jet energy
RMS.
An asymmetric window is chosen since the background is steeply
falling with increasing jet energy.

\begin{figure}\centering
\includegraphics[width=1.75in]{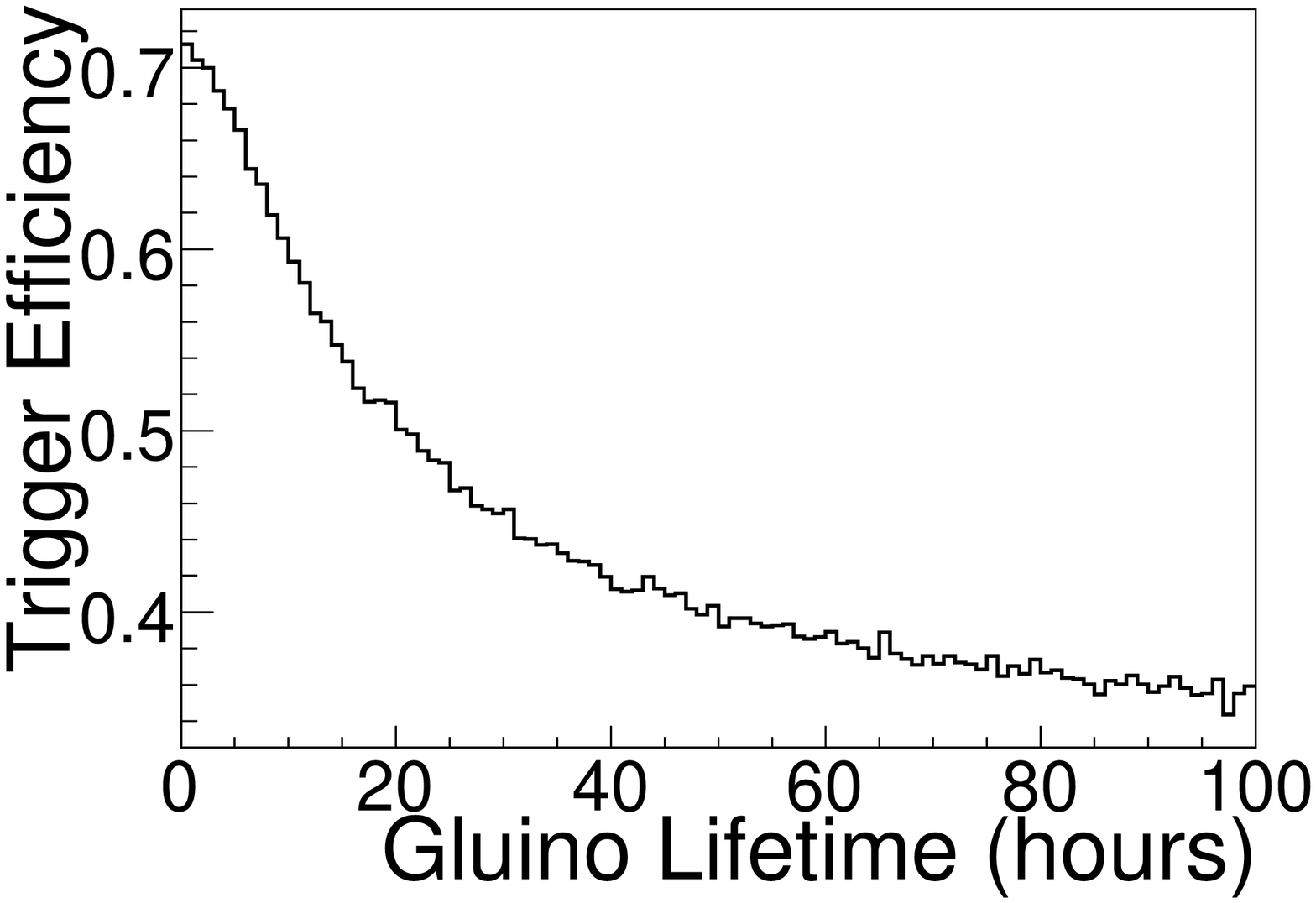}\includegraphics[width=1.75in]{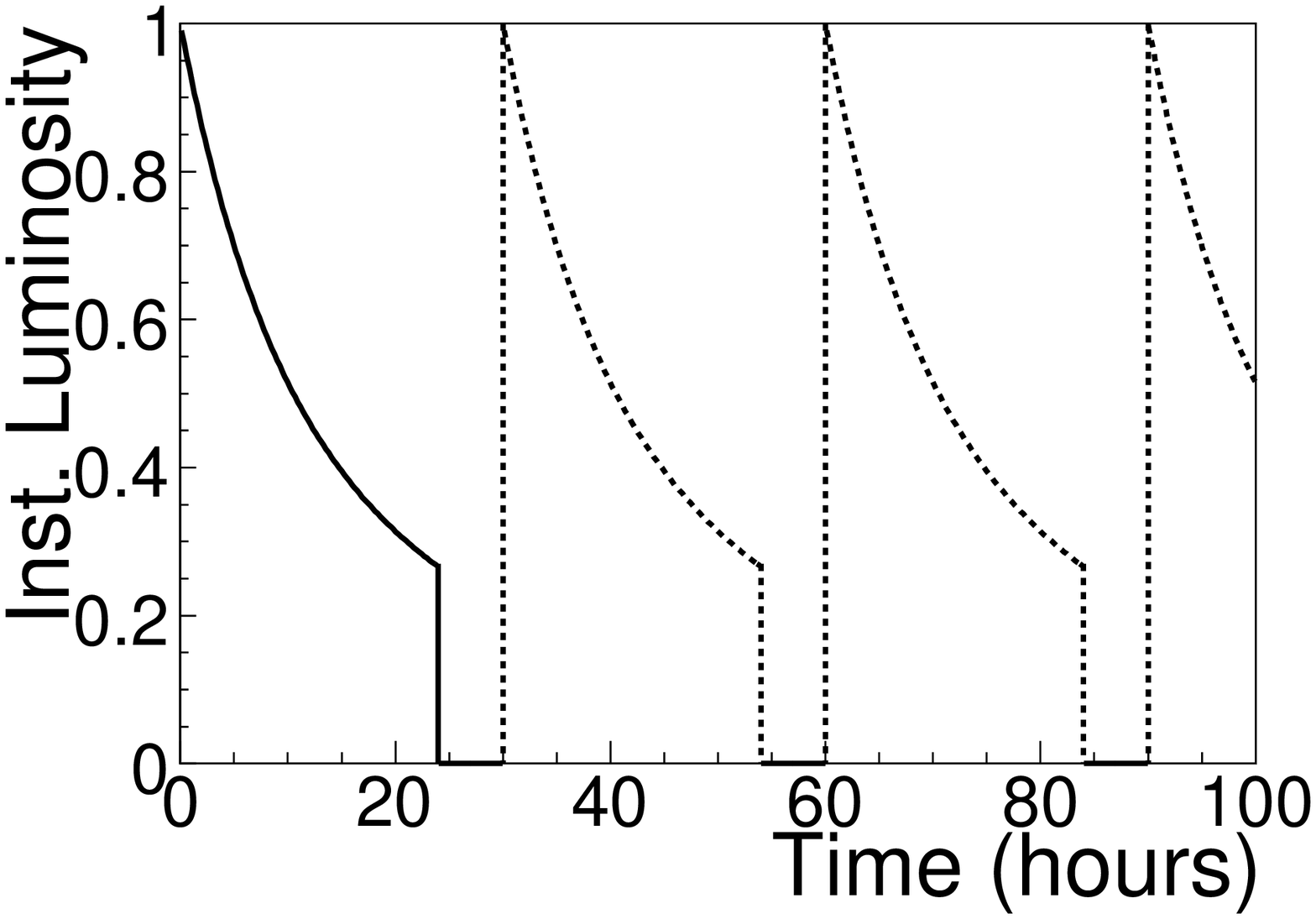}
\caption{ Left: The trigger efficiency vs.~gluino lifetime. Right:
The instantaneous luminosity profile used to model the trigger
efficiency. Dashed lines indicate a 50\% chance of the store
occurring. } \label{fig:stores}
\end{figure}

To first order, the detection efficiency for the decays of the
stopped gluino signal events can be estimated from the simulation,
but some effects are not modeled. There is a loss of efficiency at
the trigger level from the requirement of neither luminosity
scintillator array firing. If a minimum bias collision happens to
occur during the bunch crossing when the gluino decays, a luminosity
scintillator array may fire. The fraction of the time this occurs
has been measured using cosmic-muon events triggered on a jet-only
trigger with high threshold. The efficiency of the luminosity
scintillator array trigger requirement, averaged over the data set,
is 75\%. The probability to have minimum bias interactions during a
given crossing is Poisson distributed, with a mean
proportional to the instantaneous luminosity, approximately 20e30
$\text{cm}^{-1}\text{s}^{-1}$ on average for this data set.
A detailed model of the trigger efficiency is made as a function of
the gluino lifetime, for lifetimes up to 100 hours, using the
typical Tevatron store luminosity profile as input (see
Fig.~\ref{fig:stores}). Stores typically last $\sim$24 hours with a
50\% chance of another store following, 6 hours later. The current
luminosity at the time of the gluino decay, and thus the chance to
have an overlapping interaction, is accounted for. Another source of
inefficiency is that the trigger is not live all the time, but only
during the ``live super-bunches," which make up 68\% of the total
run time.

The uncertainties from all sources which affect the signal
acceptance are added in quadrature, totaling (20--25)\%. They
include the modeling of the out-of-time jet response (12\%), the
data/simulation jet energy scale (9\%), the $\eta$ and radial
distributions of stopped gluinos [(7--9)\%], other geometrical or
kinematic acceptances (5\%), and trigger efficiency [(5--15)\%].

\begin{table} \centering
\caption{The data, background, signal efficiency (for stopped
gluinos where $\tilde g$\rarrow$g\tilde{\chi}_1^0$), and expected
and observed cross section upper limits (at the 95\% C.L.) for each
jet energy range, for a small gluino lifetime, less than 3 hours.}
\begin{tabular}{cccccc}
  \hline
  \hline
  Energy (\gev) & Data & Bgnd. & Eff.(\%) & Exp. (pb) & Obs. (pb)\\
  \hline
~92.5--104.6 & 30 & 37$\pm$3.7 & 1.7$\pm$0.34 & 2.61 & 1.81 \\
112.4--156.6 & 39 & 40$\pm$4.0 & 4.9$\pm$0.98 & 0.94 & 0.89 \\
141.3--213.0 & 34 & 31$\pm$3.1 & 6.8$\pm$1.36 & 0.56 & 0.71 \\
168.7--270.6 & 32 & 26$\pm$2.6 & 7.2$\pm$1.44 & 0.48 & 0.75 \\
  \hline
  \hline
\end{tabular}
\label{table:cslimits}
\end{table}

Given an observed number of candidate events, an expected number of
background events, and a signal efficiency in a certain jet energy
range, we can exclude at the 95\% C.L. a calculated rate of signal
events giving jets of that energy, taking systematic uncertainties
into account using a Bayesian approach (see Table
\ref{table:cslimits}). This is a fairly model-independent result,
limiting the rate of any out-of-time mono-jet signal of a given
energy.


\begin{figure}\centering
\includegraphics[width=3.2in,height=1.8in]{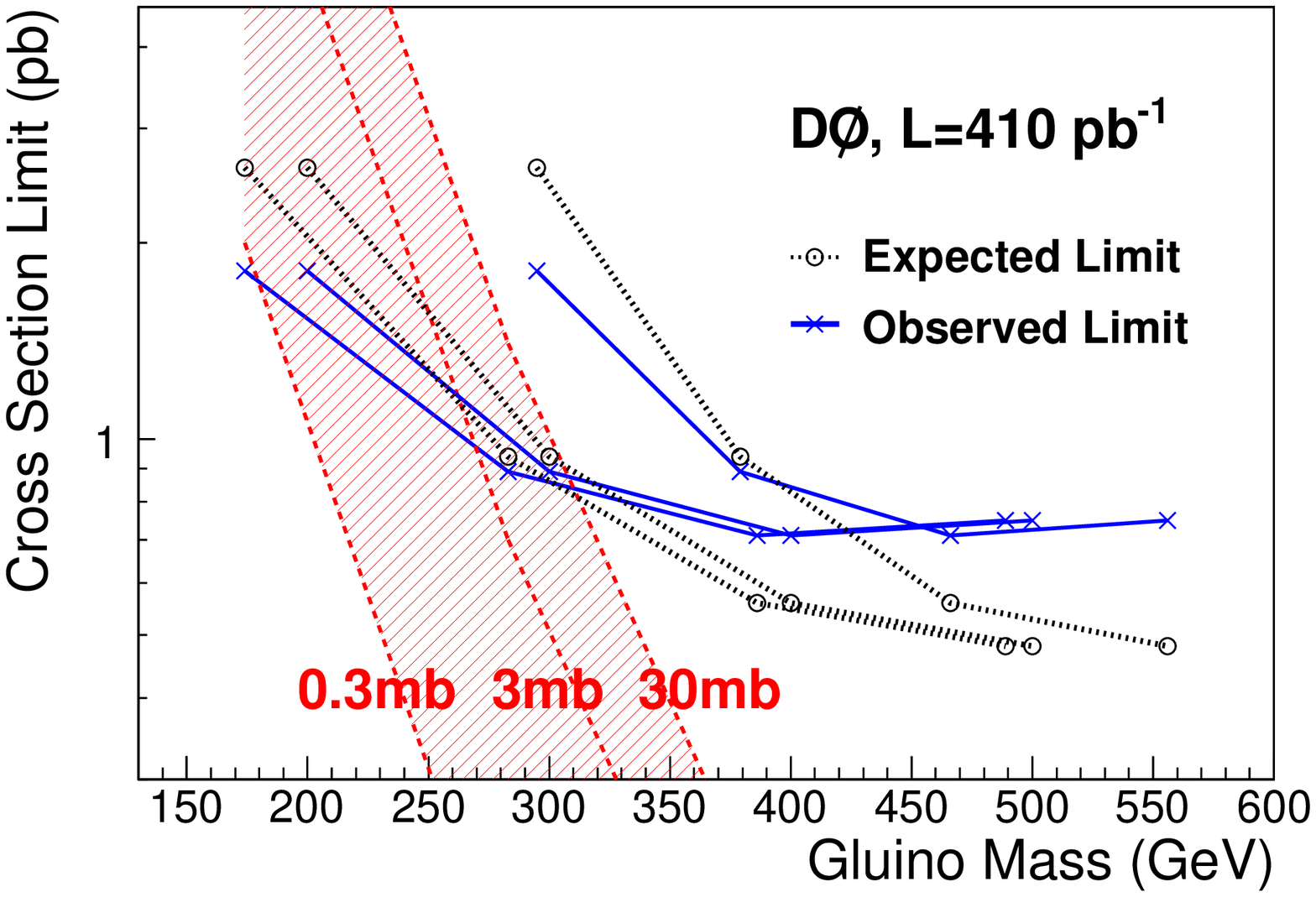}
\includegraphics[width=3.2in,height=1.8in]{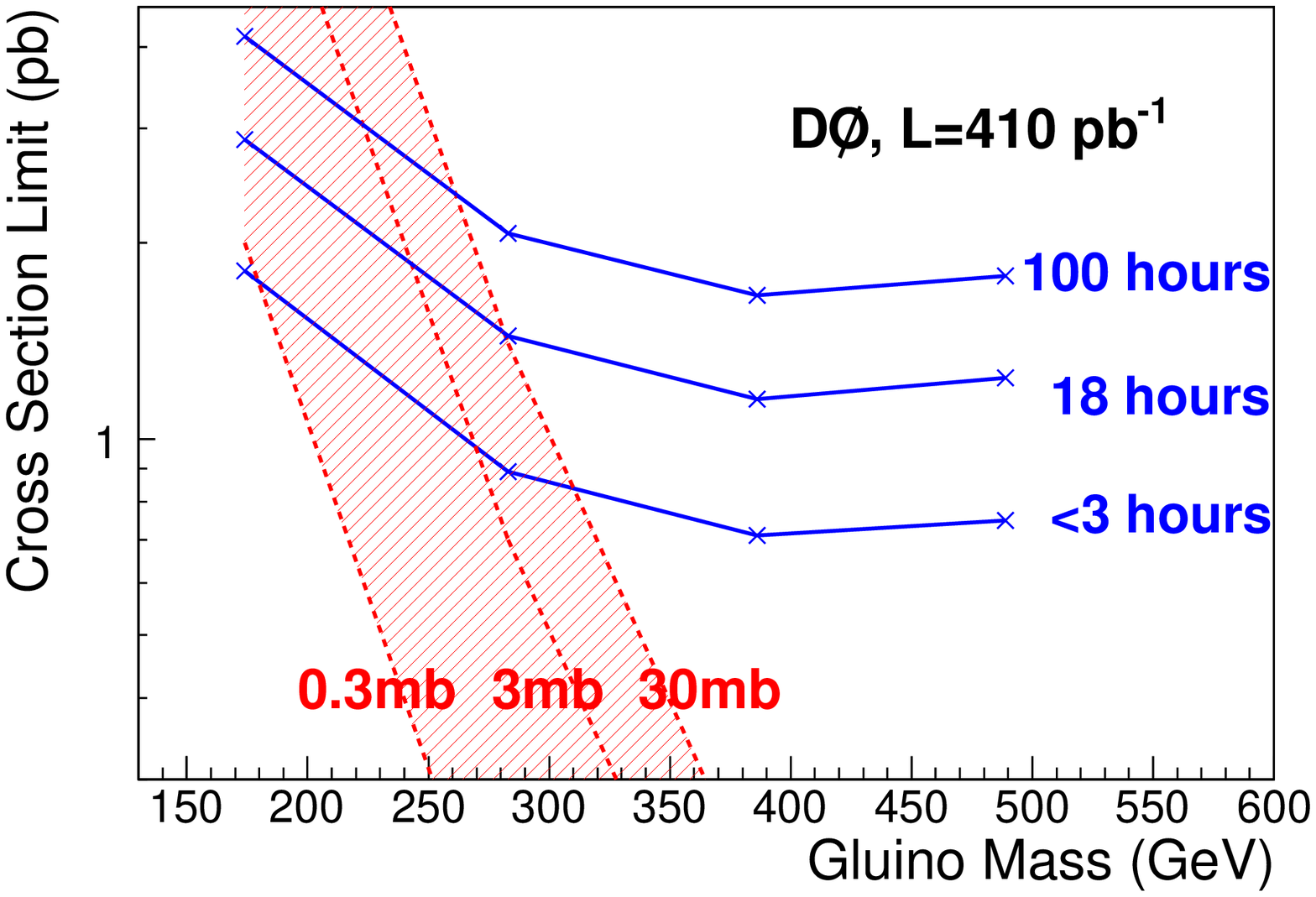}
\caption{ Top: The expected and observed upper limits on the cross
section of stopped gluinos, assuming a 100\% BR of $\tilde g$\rarrow
g$\tilde{\chi}_1^0$ and a small gluino lifetime (\lt 3 hours), for
three choices of the $\tilde{\chi}_1^0$ mass: 50, 90 and 200 \gev,
from left to right. Bottom: The upper limits observed on the cross
section of stopped gluinos, for various assumptions of the gluino
lifetime, for a $\tilde{\chi}_1^0$ mass of 50 \gev. Also shown are
the theoretical stopped gluino cross sections (dashed lines, shaded
area), from Ref.~\cite{Arvanitaki:2005nq}, for
the range of assumed conversion cross sections. }
\label{fig:limitsall}
\end{figure}

From the relation between the gluino and $\tilde{\chi}_1^0$ masses
and the observed jet energy,
results can be translated from the generated set of signal samples
to any other set of ($M_{\tilde g}$,$M_{\tilde{\chi}_1^0}$) which
would give the same jet energy. We can therefore place upper limits
on the stopped gluino cross section vs.~the gluino mass, for an
assumed $\tilde{\chi}_1^0$ mass, assuming a 100\% branching fraction
for $\tilde g$\rarrow g$\tilde{\chi}_1^0$. These can be compared
with the predicted cross sections for stopped gluinos (which include
its production rate and its probability to stop) taken from
Ref.~\cite{Arvanitaki:2005nq}. Three curves are drawn to represent
the large theory uncertainty, resulting from the variation of the
neutral to charged R-hadron conversion cross section used: 0.3, 3,
and 30 mb. Fig.~\ref{fig:limitsall} (top) shows these upper limits
for $\tilde{\chi}_1^0$ masses of 50, 90, and 200 \gev, for a small
gluino lifetime, less than 3 hours. If the gluino lifetime is
greater than 3 hours, the average efficiency of the trigger degrades
because signal events are not recorded between accelerator stores,
and the limits become weaker, as shown in Fig.~\ref{fig:limitsall}
(bottom).

This is the first search for exotic, out-of-time hadronic energy
deposits at a high-energy collider. The results from 410~\ipb\ of
Tevatron data are able to exclude a cross section of $\sim$1 pb for
gluinos stopping in the D0 calorimeter and later decaying into a
gluon and neutralino. For a $\tilde{\chi}_1^0$ mass of 50 GeV, we
are able to exclude $M_{\tilde g}$\lt 270 \gev, assuming a 100\%
branching fraction for $\tilde g$\rarrow g$\tilde{\chi}_1^0$, a
gluino lifetime less than 3 hours, and a neutral to charged R-hadron
conversion cross section of 3 mb.

Thanks to Jay Wacker
for very helpful inputs and discussions.
\input acknowledgement_paragraph_r2.tex   

\end{document}

%% file: definitions.tex
%
%
\newcommand{\comm}[1]{\mbox{\mbox{\textup{#1}}}}
\newcommand{\subs}[1]{\mbox{\scriptstyle \mathit{#1}}}
\newcommand{\subss}[1]{\mbox{\scriptscriptstyle \mathit{#1}}}
\newcommand{\Frac}[2]{\mbox{\frac{\displaystyle{#1}}{\displaystyle{#2}}}}
\newcommand{\LS}[1]{\mbox{_{\scriptstyle \mathit{#1}}}}
\newcommand{\US}[1]{\mbox{^{\scriptstyle \mathit{#1}}}}
\def\gsim{\mathrel{\rlap{\raise.4ex\hbox{$>$}} {\lower.6ex\hbox{$\sim$}}}}
\def\lsim{\mathrel{\rlap{\raise.4ex\hbox{$<$}} {\lower.6ex\hbox{$\sim$}}}}
\renewcommand{\arraystretch}{1.3}
\newcommand{\Edep}{\mbox{E_{\mathit{dep}}}}
\newcommand{\Ebeam}{\mbox{E_{\mathit{beam}}}}
\newcommand{\Exrc}{\mbox{E_{\mathit{rec}}^{\mathit{exp}}}}
\newcommand{\Emrc}{\mbox{E_{\mathit{rec}}^{\mathit{sim}}}}
\newcommand{\Evis}{\mbox{E_{\mathit{vis}}}}
\newcommand{\Edepi}{\mbox{E_{\mathit{dep,i}}}}
\newcommand{\Evisi}{\mbox{E_{\mathit{vis,i}}}}
\newcommand{\Exrci}{\mbox{E_{\mathit{rec,i}}^{\mathit{exp}}}}
\newcommand{\Emrci}{\mbox{E_{\mathit{rec,i}}^{\mathit{sim}}}}
\newcommand{\Etmis}{\mbox{E_{\mathit{t,miss}}}}
%
%
\newcommand{\lt}{\mbox{$<$}}
\newcommand{\gt}{\mbox{$>$}}
\newcommand{\lte}{\mbox{$<=$}}
\newcommand{\gte}{\mbox{$>=$}}

\newcommand{\xa}{\mbox{$x_{a}$}}
\newcommand{\xb}{\mbox{$x_{b}$}}
\newcommand{\xp}{\mbox{$x_{p}$}}
\newcommand{\xpb}{\mbox{$x_{\bar{p}}$}}

\newcommand{\alphas}{\mbox{$\alpha_{s}$}}
\newcommand{\partt}{\mbox{$\partial^{t}$}}
\newcommand{\partd}{\mbox{$\partial_{t}$}}
\newcommand{\parmut}{\mbox{$\partial^{\mu}$}}
\newcommand{\parmud}{\mbox{$\partial_{\mu}$}}
\newcommand{\parnut}{\mbox{$\partial^{\nu}$}}
\newcommand{\parnud}{\mbox{$\partial_{\nu}$}}
\newcommand{\Amut}{\mbox{$A^{\mu}$}}
\newcommand{\Amud}{\mbox{$A_{\mu}$}}
\newcommand{\Anut}{\mbox{$A^{\nu}$}}
\newcommand{\Anud}{\mbox{$A_{\nu}$}}
\newcommand{\Fmunut}{\mbox{$F^{\mu\nu}$}}
\newcommand{\Fmunud}{\mbox{$F_{\mu\nu}$}}
\newcommand{\Gmut}{\mbox{$\gamma^{\mu}$}}
\newcommand{\Gmud}{\mbox{$\gamma_{\mu}$}}
\newcommand{\Gnut}{\mbox{$\gamma^{\nu}$}}
\newcommand{\Gnud}{\mbox{$\gamma_{\nu}$}}
\newcommand{\pvecsq}{\mbox{$\vec{p}^{\:2}$}}
\newcommand{\bigsum}{\mbox{$\displaystyle{\sum}$}}
%
\newcommand{\Lzero}{Level \O}
\newcommand{\Lone}{Level $1$}
\newcommand{\Ltwo}{Level $2$}
\newcommand{\Lhalf}{Level $1.5$}
%
\newcommand{\bzero}{\mbox{\comm{B\O}}}
\newcommand{\dzero}{\mbox{D\O}}
\newcommand{\dzerosm}{\mbox{\comm{$\scriptsize{D\O}$}}}
\newcommand{\runb}{Run~$1$B}
\newcommand{\runa}{Run~$1$A}
\newcommand{\runone}{Run~$1$}
\newcommand{\runtwo}{Run~$2$}
\newcommand{\dzpjet}{\textsc{D{\O}Pjet}}
\newcommand{\y}{\mbox{$y$}}
\newcommand{\z}{\mbox{$z$}}
\newcommand{\px}{\mbox{$p_{x}$}}
\newcommand{\py}{\mbox{$p_{y}$}}
\newcommand{\pz}{\mbox{$p_{z}$}}
\newcommand{\ex}{\mbox{$E_{x}$}}
\newcommand{\ey}{\mbox{$E_{y}$}}
\newcommand{\ez}{\mbox{$E_{z}$}}
\newcommand{\et}{\mbox{$E_{T}$}}
\newcommand{\etprime}{\mbox{$E_{T}^{\prime}$}}
\newcommand{\etone}{\mbox{$E_{T}^{\mathrm{1}}$}}
\newcommand{\ettwo}{\mbox{$E_{T}^{\mathrm{2}}$}}
\newcommand{\etlj}{\mbox{$E_{T}^{\subs{lj}}$}}
\newcommand{\etmax}{\mbox{$E_{T}^{max}$}}
\newcommand{\etcand}{\mbox{$E_{T}^{\subs{cand}}$}}
\newcommand{\etup}{\mbox{$E_{T}^{\subs{up}}$}}
\newcommand{\etdown}{\mbox{$E_{T}^{\subs{down}}$}}
\newcommand{\jet}{\mbox{$E_{T}^{\subs{jet}}$}}
\newcommand{\cet}{\mbox{$E_{T}^{\subs{cell}}$}}
\newcommand{\jetvec}{\mbox{$\vec{E}_{T}^{\subs{jet}}$}}
\newcommand{\cetvec}{\mbox{$\vec{E}_{T}^{\subs{cell}}$}}
\newcommand{\jevec}{\mbox{$\vec{E}^{\subs{jet}}$}}
\newcommand{\cevec}{\mbox{$\vec{E}^{\subs{cell}}$}}
\newcommand{\etfr}{\mbox{$f_{E_{T}}$}}
\newcommand{\aveet}{\mbox{$\langle\et\rangle$}}
\newcommand{\nj}{\mbox{$n_{j}$}}
\newcommand{\ptrel}{\mbox{$p_{T}^{rel}$}}
\newcommand{\etad}{\mbox{$\eta_{d}$}}           
\newcommand{\peta}{\mbox{$\eta$}}               
\newcommand{\aeta}{\mbox{$|\eta|$}}             
\newcommand{\ifb}{fb$^{-1}$}
\newcommand{\ipb}{pb$^{-1}$}
\newcommand{\inb}{nb$^{-1}$}
\newcommand{\met}{\mbox{${\hbox{$E$\kern-0.63em\lower-.18ex\hbox{/}}}_{T}$}}
\newcommand{\metvec}{\mbox{${\hbox{$\vec{E}$\kern-0.63em\lower-.18ex\hbox{/}}}_{T}\,$}}
\newcommand{\metx}{\mbox{${\hbox{$E$\kern-0.63em\lower-.18ex\hbox{/}}}_{x}\,$}}
\newcommand{\mety}{\mbox{${\hbox{$E$\kern-0.63em\lower-.18ex\hbox{/}}}_{y}\,$}}
\newcommand{\het}{\mbox{$\vec{\mathcal{H}}_{T}$}}
\newcommand{\hetsc}{\mbox{$\mathcal{H}_{T}$}}
\newcommand{\zvrt}{\mbox{$Z$}}
\newcommand{\zcut}{\mbox{$|\zvrt| < 50$}}
\newcommand{\mtwo}{\mbox{$\mathcal{M}_{2}$}}
\newcommand{\mthree}{\mbox{$\mathcal{M}_{3}$}}
\newcommand{\mfour}{\mbox{$\mathcal{M}_{4}$}}
\newcommand{\msix}{\mbox{$\mathcal{M}_{6}$}}
\newcommand{\mn}{\mbox{$\mathcal{M}_{n}$}}
\newcommand{\R}{\mbox{$R_{\subss{MTE}}$}}
\newcommand{\invR}{\mbox{$1/R_{\subss{MTE}}$}}
\newcommand{\eemf}{\mbox{$\varepsilon_{\subss{EMF}}$}}
\newcommand{\echf}{\mbox{$\varepsilon_{\subss{CHF}}$}}
\newcommand{\ehcf}{\mbox{$\varepsilon_{\subss{HCF}}$}}
\newcommand{\eglob}{$\mbox{\varepsilon_{\subs{glob}}$}}
\newcommand{\emte}{\mbox{$\varepsilon_{\subss{MTE}}$}}
\newcommand{\ezvrt}{\mbox{$\varepsilon_{\subss{Z}}$}}
\newcommand{\etot}{\mbox{$\varepsilon_{\subs{tot}}$}}
\newcommand{\Bprime}{\mbox{$\comm{B}^{\prime}$}}
\newcommand{\Nsurv}{\mbox{$N_{\subs{surv}}$}}
\newcommand{\Nfail}{\mbox{$N_{\subs{fail}}$}}
\newcommand{\Ntot}{\mbox{$N_{\subs{tot}}$}}
\newcommand{\p}[1]{\mbox{$p_{#1}$}}
\newcommand{\ep}[1]{\mbox{$\Delta p_{#1}$}}
\newcommand{\delr}{\mbox{$\Delta R$}}
\newcommand{\deleta}{\mbox{$\Delta\eta$}}
\newcommand{\cafone}{{\sc Cafix 5.1}}
\newcommand{\caftwo}{{\sc Cafix 5.2}}
\newcommand{\delphi}{\mbox{$\Delta\varphi$}}
\newcommand{\rphi}{\mbox{$r-\varphi$}}
\newcommand{\etaphi}{\mbox{$\eta-\varphi$}}
\newcommand{\etatphi}{\mbox{$\eta\times\varphi$}}
\newcommand{\Rjet}{\mbox{$R_{jet}$}}
\newcommand{\jphi}{\mbox{$\varphi_{\subs{jet}}$}}
\newcommand{\gphi}{\mbox{$\varphi_{\subs{\gamma}}$}}
\newcommand{\ceta}{\mbox{$\eta^{\subs{cell}}$}}
\newcommand{\cphi}{\mbox{$\phi^{\subs{cell}}$}}
\newcommand{\inlum}{\mbox{$\mathcal{L}$}}
\newcommand{\gm}{\mbox{$\gamma$}}
\newcommand{\Rjj}{\mbox{$\mathbf{R}_{\subs{jj}}$}}
\newcommand{\Rgj}{\mbox{$\mathbf{R}_{\subs{\gamma j}}$}}
\newcommand{\Rmathcal}{\mbox{$\mathcal{R}$}}
\newcommand{\etv}{\mbox{$\vec{E}_{T}$}}
\newcommand{\nvec}{\mbox{$\hat{\vec{n}}$}}
\newcommand{\eprime}{\mbox{$E^{\prime}$}}
\newcommand{\aveprime}{\mbox{$\bar{E}^{\prime}$}}
\newcommand{\geta}{\mbox{$\eta_{\gm}$}}
\newcommand{\jeta}{\mbox{$\eta_{\subs{jet}}$}}
\newcommand{\cjeta}{\mbox{$\eta_{\subs{jet}}^{\subss{CEN}}$}}
\newcommand{\fjeta}{\mbox{$\eta_{\subs{jet}}^{\subss{FOR}}$}}
\newcommand{\cjphi}{\mbox{$\varphi_{\subs{jet}}^{\subss{CEN}}$}}
\newcommand{\fjphi}{\mbox{$\varphi_{\subs{jet}}^{\subss{FOR}}$}}
\newcommand{\etcut}{\mbox{$E_{T}^{\subs{cut}}$}}
\newcommand{\etg}{\mbox{$E_{T}^{\gm}$}}
\newcommand{\cenet}{\mbox{$E_{T}^{\subss{CEN}}$}}
\newcommand{\foret}{\mbox{$E_{T}^{\subss{FOR}}$}}
\newcommand{\cenen}{\mbox{$E^{\subss{CEN}}$}}
\newcommand{\foren}{\mbox{$E^{\subss{FOR}}$}}
\newcommand{\ejtptc}{\mbox{$E^{\subs{ptcl}}_{\subs{jet}}$}}
\newcommand{\ejtmes}{\mbox{$E^{\subs{meas}}_{\subs{jet}}$}}
\newcommand{\AIDA}{{\sc AIDA}}
\newcommand{\RECO}{{\sc Reco}}
\newcommand{\PYTHIA}{{\sc Pythia}}
\newcommand{\HERWIG}{{\sc Herwig}}
\newcommand{\JETRAD}{{\sc Jetrad}}
\newcommand{\CTone}{\mbox{$|\eta|<0.4}$}
\newcommand{\CTtwo}{\mbox{$0.4\leq|\eta|<0.8$}}
\newcommand{\ICone}{\mbox{$0.8\leq|\eta|<1.2$}}
\newcommand{\ICtwo}{\mbox{$1.2\leq|\eta|<1.6$}}
\newcommand{\FWone}{\mbox{$1.6\leq|\eta|<2.0$}}
\newcommand{\FWtwo}{\mbox{$2.0\leq|\eta|<2.5$}}
\newcommand{\FWthr}{\mbox{$2.5\leq|\eta|<3.0$}}
\newcommand{\LCTone}{\mbox{$|\eta|<0.5$}}
\newcommand{\LCTtwo}{\mbox{$0.5\leq|\eta|<1.0$}}
\newcommand{\LICone}{\mbox{$1.0\leq|\eta|<1.5$}}
\newcommand{\LICtwo}{\mbox{$1.5\leq|\eta|<2.0$}}
\newcommand{\LFWone}{\mbox{$2.0\leq|\eta|<3.0$}}
\newcommand{\CSone}{\mbox{$|\eta|<0.5$}}
\newcommand{\CStwo}{\mbox{$0.5\leq|\eta|<1.0$}}
\newcommand{\CSthr}{\mbox{$1.0\leq|\eta|<1.5$}}
\newcommand{\CSfou}{\mbox{$1.5\leq|\eta|<2.0$}}
\newcommand{\CSfiv}{\mbox{$2.0\leq|\eta|<3.0$}}
\newcommand{\sigA}{\mbox{$\sigma_{\subss{\!A}}$}}
\newcommand{\sigASS}{\mbox{$\sigma_{\subss{A}}^{\subss{SS}}$}}
\newcommand{\sigAOS}{\mbox{$\sigma_{\subss{A}}^{\subss{OS}}$}}
\newcommand{\sigZ}{\mbox{$\sigma_{\subss{Z}}$}}
\newcommand{\sige}{\mbox{$\sigma_{\subss{E}}$}}
\newcommand{\siget}{\mbox{$\sigma_{\subss{\et}}$}}
\newcommand{\sigetone}{\mbox{$\sigma_{\subs{\etone}}$}}
\newcommand{\sigettwo}{\mbox{$\sigma_{\subs{\ettwo}}$}}
\newcommand{\rcal}{\mbox{$R_{\subs{cal}}$}}
\newcommand{\zcal}{\mbox{$Z_{\subs{cal}}$}}
\newcommand{\Runf}{\mbox{$R_{\subs{unf}}$}}
\newcommand{\Rsep}{\mbox{$\mathcal{R}_{sep}$}}
\newcommand{\etal}{{\it et al.}}
\newcommand{\ppbar}{\mbox{$p\overline{p}$}}
\newcommand{\pp}{\mbox{$pp$}}
\newcommand{\qqbar}{\mbox{$q\overline{q}$}}
\newcommand{\ccbar}{\mbox{$c\overline{c}$}}
\newcommand{\bbbar}{\mbox{$b\overline{b}$}}
\newcommand{\ttbar}{\mbox{$t\overline{t}$}}

\newcommand{\bbj}{\mbox{$b\overline{b}j$}}
\newcommand{\bbjj}{\mbox{$b\overline{b}jj$}}
\newcommand{\ccjj}{\mbox{$c\overline{c}jj$}}

\newcommand{\bb}{\mbox{$b\overline{b}j(j)$}}
\newcommand{\cc}{\mbox{$c\overline{c}j(j)$}}

\newcommand{\hboson}{\mbox{$\mathit{h}$}}
\newcommand{\Hboson}{\mbox{$\mathit{H}$}}
\newcommand{\Aboson}{\mbox{$\mathit{A}$}}
\newcommand{\zboson}{\mbox{$\mathit{Z}$}}
\newcommand{\zb}{\mbox{$\mathit{Zb}$}}
\newcommand{\bh}{\mbox{$\mathit{bh}$}}
\newcommand{\btag}{\mbox{$\mathit{b}$}}

\newcommand{\hsm}{\mbox{$h_{SM}$}}
\newcommand{\hmssm}{\mbox{$h_{MSSM}$}}

\newcommand{\prot}{\mbox{$p$}}
\newcommand{\pbar}{\mbox{$\overline{p}$}}
\newcommand{\pt}{\mbox{$p_{T}$}}
\newcommand{\xnot}{\mbox{$X_{0}$}}
\newcommand{\Znot}{\mbox{$Z^{0}$}}
\newcommand{\Wpm}{\mbox{$W^{\pm}$}}
\newcommand{\Wplus}{\mbox{$W^{+}$}}
\newcommand{\Wminus}{\mbox{$W^{-}$}}
\newcommand{\lamb}{\mbox{$\lambda$}}
\newcommand{\nhatbf}{\mbox{$\hat{\mathbf{n}}$}}
\newcommand{\pbf}{\mbox{$\mathbf{p}$}}
\newcommand{\xbf}{\mbox{$\mathbf{x}$}}
\newcommand{\jbf}{\mbox{$\mathbf{j}$}}
\newcommand{\Ebf}{\mbox{$\mathbf{E}$}}
\newcommand{\Bbf}{\mbox{$\mathbf{B}$}}
\newcommand{\Abf}{\mbox{$\mathbf{A}$}}
\newcommand{\Rbf}{\mbox{$\mathbf{R}$}}
\newcommand{\nablabf}{\mbox{$\mathbf{\nabla}$}}
\newcommand{\rarrow}{\mbox{$\rightarrow$}}
\newcommand{\slashp}{\mbox{$\not \! p \,$}}
\newcommand{\slashk}{\mbox{$\not \! k$}}
\newcommand{\slasha}{\mbox{$\not \! a$}}
\newcommand{\slashA}{\mbox{$\! \not \! \! A$}}
\newcommand{\slashpar}{\mbox{$\! \not \! \partial$}}
\newcommand{\intdouble}{\mbox{$\int\!\!\int$}}
\newcommand{\MRSTGU}{MRSTg$\uparrow$}
\newcommand{\MRSTGD}{MRSTg$\downarrow$}
%
\newcommand{\Due}{\mbox{$D_{\mathrm{ue}}$}}
\newcommand{\Dth}{\mbox{$D_{\Theta}$}}
\newcommand{\Dof}{\mbox{$D_{\mathrm{O}}$}}
\newcommand{\zbl}{\texttt{ZERO BIAS}}
\newcommand{\mbl}{\texttt{MIN BIAS}}
\newcommand{\mbll}{\texttt{MINIMUM BIAS}}
\newcommand{\nue}{\mbox{$\nu_{e}$}}
\newcommand{\num}{\mbox{$\nu_{\mu}$}}
\newcommand{\nut}{\mbox{$\nu_{\tau}$}}
\newcommand{\mycs}{\mbox{$d^{\,2}\sigma/(d\et d\eta)$}}
\newcommand{\mycsav}{\mbox{$\langle \mycs \rangle$}}
\newcommand{\tdcs}{\mbox{$d^{\,3}\sigma/d\et d\eta_{1} d\eta_{2}$}}
\newcommand{\tdcsav}{\mbox{$\langle d^{\,3}\sigma/d\et d\eta_{1} d\eta_{2} \rangle$}}
\newcommand{\tanb}{$\tan\beta$}
\newcommand{\cotb}{$\cot\beta$}

\newcommand{\rstev}{\mbox{$\rs = \T{1.8}$}}
\newcommand{\rssps}{\mbox{$\rs = \T{0.63}$}}
\newcommand{\XX}{\mbox{$\, \times \,$}}
\newcommand{\AP}{\mbox{${\rm \bar{p}}$}}
\newcommand{\SU}{\mbox{$<\! |S|^2 \!>$}}
\newcommand{\ET}{\mbox{$E_{T}$}}
\newcommand{\HT}{\mbox{$S_{{\rm {\sl T}}}$} }
\newcommand{\PT}{\mbox{$p_{t}$}}
\newcommand{\DP}{\mbox{$\Delta\phi$}}
\newcommand{\DR}{\mbox{$\Delta R$}}
\newcommand{\DE}{\mbox{$\Delta\eta$}}
\newcommand{\DEP}{\mbox{$\Delta\eta_{c}$}}
\newcommand{\PH}{\mbox{$\phi$}}
\newcommand{\EA}{\mbox{$\eta$} }
\newcommand{\EAJ}{\mbox{\EA(jet)}}
\newcommand{\AEA}{\mbox{$|\eta|$}}
\newcommand{\Ge}[1]{\mbox{#1 GeV}}
\newcommand{\T}[1]{\mbox{#1 TeV}}
\newcommand{\x}{\cdot}
\newcommand{\ra}{\rightarrow}
\def\D0{D\O}
\def\ETmiss{{\rm {\mbox{$E\kern-0.57em\raise0.19ex\hbox{/}_{T}$}}}}
\newcommand{\mb}{\mbox{mb}}
\newcommand{\nb}{\mbox{nb}}
\newcommand{\rs}{\mbox{$\sqrt{\rm {\sl s}}$}}
\newcommand{\fdel}{\mbox{$f(\DEP)$}}
\newcommand{\fdele}{\mbox{$f(\DEP)^{exp}$}}
\newcommand{\fgap}{\mbox{$f(\DEP\! \geq \!3)$}}
\newcommand{\fgape}{\mbox{$f(\DEP\! \geq \!3)^{exp}$}}
\newcommand{\fpyt}{\mbox{$f(\DEP\!>\!2)$}}
\newcommand{\delth}{\mbox{$\DEP\! \geq \!3$}}
\newcommand{\uplim}{\mbox{$1.1\!\times\!10^{-2}$}}
\def\simge
{\mathrel{\rlap{\raise 0.53ex \hbox{$>$}}{\lower 0.53ex \hbox{$\sim$}}}}
\def\simle
{\mathrel{\rlap{\raise 0.53ex \hbox{$<$}}{\lower 0.53ex \hbox{$\sim$}}}}
\newcommand{\pbarp}{\mbox{$p\bar{p}$}}
\def\ETmiss{\mbox{${\hbox{$E$\kern-0.5em\lower-.1ex\hbox{/}\kern+0.15em}}_{\rm T}$}}
\def\Et{\mbox{$E_{T}$}}
\newcommand{\modeta}{\mid \!\! \eta \!\! \mid}
\def\gevcc{GeV/c$^2$}                   
\def\gevc{GeV/c}                        
\def\gev{GeV}                           
\newcommand{\als}{\mbox{${\alpha_{{\rm s}}}$}}
\def\1960{$\sqrt{s}=1.96$ TeV}
\def\etI{E_{T_1}}
\def\etII{E_{T_2}}
\def\itaI{\eta_1}
\def\itaII{\eta_2}
\def\deta{\Delta\eta}
\def\etab{\bar{\eta}}
\def\xq{($x_1$,$x_2$,$Q^2$)}
\def\xx{($x_1$,$x_2$)}
\def\rap{pseudorapidity}
\def\as{\alpha_s}
\def\ap{\alpha_{\rm BFKL}}
\def\apb{\alpha_{{\rm BFKL}_{bin}}}
\def\cm{c.m.}

%% file: list_of_authors_r2.tex
%
\author{                                                                      
V.M.~Abazov,$^{35}$                                                           
B.~Abbott,$^{75}$                                                             
M.~Abolins,$^{65}$                                                            
B.S.~Acharya,$^{28}$                                                          
M.~Adams,$^{51}$                                                              
T.~Adams,$^{49}$                                                              
E.~Aguilo,$^{5}$                                                              
S.H.~Ahn,$^{30}$                                                              
M.~Ahsan,$^{59}$                                                              
G.D.~Alexeev,$^{35}$
% ...
\\                                                                            
\vskip 0.30cm                                                                 
\centerline{(D\O\ Collaboration)}                                             
\vskip 0.30cm                                                                 
}                                                                             
\affiliation{                                                                 
\centerline{$^{1}$Universidad de Buenos Aires, Buenos Aires, Argentina}       
\centerline{$^{2}$LAFEX, Centro Brasileiro de Pesquisas F{\'\i}sicas,         
                  Rio de Janeiro, Brazil}                                     
\centerline{$^{3}$Universidade do Estado do Rio de Janeiro,                   
                  Rio de Janeiro, Brazil}                                     
\centerline{$^{4}$Instituto de F\'{\i}sica Te\'orica, Universidade            
                  Estadual Paulista, S\~ao Paulo, Brazil}                     
\centerline{$^{5}$University of Alberta, Edmonton, Alberta, Canada,           
                  Simon Fraser University, Burnaby, British Columbia, Canada,}
\centerline{York University, Toronto, Ontario, Canada, and                    
                  McGill University, Montreal, Quebec, Canada}                
\centerline{$^{6}$University of Science and Technology of China, Hefei,       
                  People's Republic of China}                                 
\centerline{$^{7}$Universidad de los Andes, Bogot\'{a}, Colombia}             
\centerline{$^{8}$Center for Particle Physics, Charles University,            
                  Prague, Czech Republic}                                     
\centerline{$^{9}$Czech Technical University, Prague, Czech Republic}         
\centerline{$^{10}$Center for Particle Physics, Institute of Physics,         
                   Academy of Sciences of the Czech Republic,                 
                   Prague, Czech Republic}                                    
\centerline{$^{11}$Universidad San Francisco de Quito, Quito, Ecuador}        
\centerline{$^{12}$Laboratoire de Physique Corpusculaire, IN2P3-CNRS,         
                   Universit\'e Blaise Pascal, Clermont-Ferrand, France}      
\centerline{$^{13}$Laboratoire de Physique Subatomique et de Cosmologie,      
                   IN2P3-CNRS, Universite de Grenoble 1, Grenoble, France}    
\centerline{$^{14}$CPPM, IN2P3-CNRS, Universit\'e de la M\'editerran\'ee,     
                   Marseille, France}                                         
\centerline{$^{15}$Laboratoire de l'Acc\'el\'erateur Lin\'eaire,              
                   IN2P3-CNRS et Universit\'e Paris-Sud, Orsay, France}       
\centerline{$^{16}$LPNHE, IN2P3-CNRS, Universit\'es Paris VI and VII,         
                   Paris, France}                                             
\centerline{$^{17}$DAPNIA/Service de Physique des Particules, CEA, Saclay,    
                   France}                                                    
\centerline{$^{18}$IPHC, Universit\'e Louis Pasteur et Universit\'e           
                   de Haute Alsace, CNRS, IN2P3, Strasbourg, France}          
\centerline{$^{19}$IPNL, Universit\'e Lyon 1, CNRS/IN2P3, Villeurbanne, France
                   and Universit\'e de Lyon, Lyon, France}                    
\centerline{$^{20}$III. Physikalisches Institut A, RWTH Aachen,               
                   Aachen, Germany}                                           
\centerline{$^{21}$Physikalisches Institut, Universit{\"a}t Bonn,             
                   Bonn, Germany}                                             
\centerline{$^{22}$Physikalisches Institut, Universit{\"a}t Freiburg,         
                   Freiburg, Germany}                                         
\centerline{$^{23}$Institut f{\"u}r Physik, Universit{\"a}t Mainz,            
                   Mainz, Germany}                                            
\centerline{$^{24}$Ludwig-Maximilians-Universit{\"a}t M{\"u}nchen,            
                   M{\"u}nchen, Germany}                                      
\centerline{$^{25}$Fachbereich Physik, University of Wuppertal,               
                   Wuppertal, Germany}                                        
\centerline{$^{26}$Panjab University, Chandigarh, India}                      
\centerline{$^{27}$Delhi University, Delhi, India}                            
\centerline{$^{28}$Tata Institute of Fundamental Research, Mumbai, India}     
\centerline{$^{29}$University College Dublin, Dublin, Ireland}                
\centerline{$^{30}$Korea Detector Laboratory, Korea University,               
                   Seoul, Korea}                                              
\centerline{$^{31}$SungKyunKwan University, Suwon, Korea}                     
\centerline{$^{32}$CINVESTAV, Mexico City, Mexico}                            
\centerline{$^{33}$FOM-Institute NIKHEF and University of                     
                   Amsterdam/NIKHEF, Amsterdam, The Netherlands}              
\centerline{$^{34}$Radboud University Nijmegen/NIKHEF, Nijmegen, The          
                  Netherlands}                                                
\centerline{$^{35}$Joint Institute for Nuclear Research, Dubna, Russia}       
\centerline{$^{36}$Institute for Theoretical and Experimental Physics,        
                   Moscow, Russia}                                            
\centerline{$^{37}$Moscow State University, Moscow, Russia}                   
\centerline{$^{38}$Institute for High Energy Physics, Protvino, Russia}       
\centerline{$^{39}$Petersburg Nuclear Physics Institute,                      
                   St. Petersburg, Russia}                                    
\centerline{$^{40}$Lund University, Lund, Sweden, Royal Institute of          
                   Technology and Stockholm University, Stockholm,            
                   Sweden, and}                                               
\centerline{Uppsala University, Uppsala, Sweden}                              
\centerline{$^{41}$Physik Institut der Universit{\"a}t Z{\"u}rich,            
                   Z{\"u}rich, Switzerland}                                   
\centerline{$^{42}$Lancaster University, Lancaster, United Kingdom}           
\centerline{$^{43}$Imperial College, London, United Kingdom}                  
\centerline{$^{44}$University of Manchester, Manchester, United Kingdom}      
\centerline{$^{45}$University of Arizona, Tucson, Arizona 85721, USA}         
\centerline{$^{46}$Lawrence Berkeley National Laboratory and University of    
                   California, Berkeley, California 94720, USA}               
\centerline{$^{47}$California State University, Fresno, California 93740, USA}
\centerline{$^{48}$University of California, Riverside, California 92521, USA}
\centerline{$^{49}$Florida State University, Tallahassee, Florida 32306, USA} 
\centerline{$^{50}$Fermi National Accelerator Laboratory,                     
            Batavia, Illinois 60510, USA}                                     
\centerline{$^{51}$University of Illinois at Chicago,                         
            Chicago, Illinois 60607, USA}                                     
\centerline{$^{52}$Northern Illinois University, DeKalb, Illinois 60115, USA} 
\centerline{$^{53}$Northwestern University, Evanston, Illinois 60208, USA}    
\centerline{$^{54}$Indiana University, Bloomington, Indiana 47405, USA}       
\centerline{$^{55}$University of Notre Dame, Notre Dame, Indiana 46556, USA}  
\centerline{$^{56}$Purdue University Calumet, Hammond, Indiana 46323, USA}    
\centerline{$^{57}$Iowa State University, Ames, Iowa 50011, USA}              
\centerline{$^{58}$University of Kansas, Lawrence, Kansas 66045, USA}         
\centerline{$^{59}$Kansas State University, Manhattan, Kansas 66506, USA}     
\centerline{$^{60}$Louisiana Tech University, Ruston, Louisiana 71272, USA}   
\centerline{$^{61}$University of Maryland, College Park, Maryland 20742, USA} 
\centerline{$^{62}$Boston University, Boston, Massachusetts 02215, USA}       
\centerline{$^{63}$Northeastern University, Boston, Massachusetts 02115, USA} 
\centerline{$^{64}$University of Michigan, Ann Arbor, Michigan 48109, USA}    
\centerline{$^{65}$Michigan State University,                                 
            East Lansing, Michigan 48824, USA}                                
\centerline{$^{66}$University of Mississippi,                                 
            University, Mississippi 38677, USA}                               
\centerline{$^{67}$University of Nebraska, Lincoln, Nebraska 68588, USA}      
\centerline{$^{68}$Princeton University, Princeton, New Jersey 08544, USA}    
\centerline{$^{69}$State University of New York, Buffalo, New York 14260, USA}
\centerline{$^{70}$Columbia University, New York, New York 10027, USA}        
\centerline{$^{71}$University of Rochester, Rochester, New York 14627, USA}   
\centerline{$^{72}$State University of New York,                              
            Stony Brook, New York 11794, USA}                                 
\centerline{$^{73}$Brookhaven National Laboratory, Upton, New York 11973, USA}
\centerline{$^{74}$Langston University, Langston, Oklahoma 73050, USA}        
\centerline{$^{75}$University of Oklahoma, Norman, Oklahoma 73019, USA}       
\centerline{$^{76}$Oklahoma State University, Stillwater, Oklahoma 74078, USA}
\centerline{$^{77}$Brown University, Providence, Rhode Island 02912, USA}     
\centerline{$^{78}$University of Texas, Arlington, Texas 76019, USA}          
\centerline{$^{79}$Southern Methodist University, Dallas, Texas 75275, USA}   
\centerline{$^{80}$Rice University, Houston, Texas 77005, USA}                
\centerline{$^{81}$University of Virginia, Charlottesville,                   
            Virginia 22901, USA}                                              
\centerline{$^{82}$University of Washington, Seattle, Washington 98195, USA}  
}                                                                             

%% file: acknowledgement_paragraph_r2.tex
%
We thank the staffs at Fermilab and collaborating institutions, 
and acknowledge support from the 
DOE and NSF (USA);
CEA and CNRS/IN2P3 (France);
FASI, Rosatom and RFBR (Russia);
CAPES, CNPq, FAPERJ, FAPESP and FUNDUNESP (Brazil);
DAE and DST (India);
Colciencias (Colombia);
CONACyT (Mexico);
KRF and KOSEF (Korea);
CONICET and UBACyT (Argentina);
FOM (The Netherlands);
Science and Technology Facilities Council (United Kingdom);
MSMT and GACR (Czech Republic);
CRC Program, CFI, NSERC and WestGrid Project (Canada);
BMBF and DFG (Germany);
SFI (Ireland);
The Swedish Research Council (Sweden);
CAS and CNSF (China);
Alexander von Humboldt Foundation;
and the Marie Curie Program.
%